# Exciton- and Light-induced Current in Molecular Nanojunctions

B.D. Fainberg[a,b], P. Hanggi[c], S. Kohler[c] and A. Nitzan[b]

[a]*Faculty of Sciences, Holon Institute of Technology, 52 Golomb St., Holon 58102, Israel*
[b]*School of Chemistry, Tel-AvivUniversity, Tel-Aviv 69978, Israel*
[c] *Institute for Physics, University of Augsburg, Augsburg, D-86135, Germany*

**Abstract.** We consider exciton- and light-induced current in molecular nanojunctions. Using a model comprising a two two-level sites bridge connecting free electron reservoirs we show that the exciton coupling between the sites of the molecular bridge can markedly effect the source-drain current through a molecular junction. In some cases when excited and unexcited states of the sites are coupled differently to the leads, the contribution from electron-hole excitations can exceed the Landauer elastic current and dominate the observed conduction. We have proposed an optical control method using chirped pulses for enhancing charge transfer in unbiased junctions where the bridging molecule is characterized by a strong charge-transfer transition.

## INTRODUCTION

Electron transport through molecular wires has been under intense study in the last few years [1-3]. Theoretical modeling of electron transport [2,3] starts from the wire Hamiltonian as a tight-binding model composed of $N$ sites that contain electron transfer (tunneling) interactions between nearest sites. For a molecular wire, this constitutes the so-called Huckel description where each site corresponds to one atom. Necessary conditions for finite current in this model are, first, the existence of such interactions between (quasi-)resonant states of nearest sites, and second, a biased junction. In this presentation we consider additional interactions, which enable us to remove one of these conditions for current to occur.

## COHERENT CHARGE TRANSPORT THROUGH MOLECULAR WIRES: CURRENT FROM ELECTRONIC EXCITATION IN THE WIRE

For a typical distance of 5 Å between two neighboring sites, which can be either atoms or molecules in molecular assemblies, energy-transfer interactions – excitation (deexcitation) of a site accompanied by deexcitation (excitation) of its nearest neighbor - are well-known in the exciton theory [4]. Electron transfer is a tunneling process that depends exponentially on the site-site distance, while energy transfer is associated with dipolar coupling that scales like the inverse cube of this distance, and can therefore dominate at larger distances. To the best of our knowledge, there were no previous treatments of transport in molecular wires that take into account simultaneous effects of both electron and energy transfer. Here we address this problem by using the Liouville-von Neumann equation (LNE) for the total density operator to derive an expression for the conduction of a molecular wire model that contains both electron and energy-transfer interactions, then analyze several examples with reasonable parameters. We show that the effect of exciton type interactions on electron transport through moleculars wires can be significant, sometimes even dominant, in a number of situations. In particular, the current occurs even when the above mentioned first condition is not fulfilled.

## Model

We consider a molecular wire that comprises a dimer represented by its highest occupied molecular orbitals (HOMO), $|g\rangle$, and lowest unoccupied molecular orbitals (LUMO), $|e\rangle$, positioned between two leads represented by free electron reservoirs $L$ and $R$ (Fig.1). The electron reservoirs (leads) are characterized by their electronic chemical potentials $\mu_L$ and $\mu_R$, where the difference $\mu_L - \mu_R = e\Phi$ is the imposed voltage bias. The corresponding Hamiltonian is

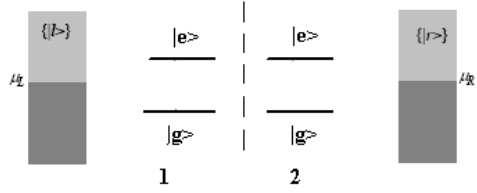

**FIGURE 1.** A model for energy-transfer induced effects in molecular conduction. The right ($R=|\{r\}\rangle$) and left ($L=|\{l\}\rangle$) manifolds represent the two metal leads characterized by electrochemical potentials $\mu_R$ and $\mu_L$ respectively. A molecular dimer is represented by its HOMOs, $|1g\rangle$ and $|2g\rangle$, and LUMOs, $|1e\rangle$ and $|2e\rangle$.

$$\hat{H} = \hat{H}_{wire} + \hat{H}_{leads} + \hat{H}_{contacts} \qquad (1)$$

where the different terms correspond to the wire, the leads ($\hat{H}_{leads} = \sum_{k \in \{L,R\}} \varepsilon_k \hat{c}_k^+ \hat{c}_k$), and the wire-lead couplings ($\hat{H}_{contacts} = \hat{V}_M + \hat{V}_N$), respectively. $\hat{V}_M = \sum_{nkf} V_{nf,k}^{(MK_n)} \hat{c}_k^+ \hat{c}_{nf} + H.c.$ describes electron transfer between the molecular bridge and the leads that gives rise to net current in the biased junction while

$\hat{V}_N = \sum_{n, k \neq k'} V_{kk'}^{(NK_n)} \hat{c}_k^+ \hat{c}_{k'} b_n^+ + H.c.$ describes energy transfer between the bridge and electron-hole excitations in the leads. Here *H.c.* denotes Hermitian conjugate, $L$ and $R$ denote the left and right leads, respectively, and $K_1 = L$, $K_2 = R$.

$$\hat{H}_{wire} = \sum_{\substack{m=1,2 \\ f=g,e}} \varepsilon_{mf} \hat{c}_{mf}^+ \hat{c}_{mf} - \sum_{f=g,e} \Delta_f (\hat{c}_{2f}^+ \hat{c}_{1f} + \hat{c}_{1f}^+ \hat{c}_{2f}) + \hbar J (b_1^+ b_2 + b_2^+ b_1) + \sum_{m=1,2} U_m N_m (N_m - 1)$$

(2)

The operators $\hat{c}_{mf}^+$ ($\hat{c}_{mf}$) create (annihilate) an electron in the orbital $|mf\rangle$, and $\varepsilon_{mf}$ denotes the respective on-site energy, $\hat{n}_{mf} = \hat{c}_{mf}^+ \hat{c}_{mf}$. The second and the third terms on the RHS of Eq. 2 describe electron and energy transfer between the sites, respectively. Since we aim at exploring blocking effects, the last term on the RHS of Eq. 2 takes account of the Coulomb repulsion on a site in the limit of large interaction strengths $U_m$ where $N_m = \sum_{f=g,e} \hat{n}_{mf}$ is the operator counting the excess electrons on the sites. The excitonic operators are equal to $b_m^+ = \hat{c}_{me}^+ \hat{c}_{mg}$. The effect of the corresponding interaction in the bridge ( $= \hbar J b_1^+ b_2 + H.c.$ ) on the charge transport properties is the subject of our discussion.

## Master Equation in the Eigenbasis of Many-electron Wire Hamiltonian

The central idea of using LNE for the computation of stationary currents is to consider $\hat{H}_{contacts}$ as a perturbation. For the total density operator $\rho$ one can obtain by standard techniques the approximate equation of motion [2,3]. The information of interest is limited only to the wire part of the density operator $\sigma(t)$, which can be obtained by defining a projection operator $P$ that projects the complete system onto the relevant (molecule) part and by tracing out the reservoir degrees of freedom ( $K = L, R$ ): $P\rho(t) = \rho_K Tr_K \rho(t) = \rho_K \otimes \hat{\sigma}(t)$ with reservoir density matrix $\rho_K$. As to $\rho_K$, we employ the grand-canonical ensemble of non-interacting electrons in the leads at temperature $T$, characterized by electrochemical potentials $\mu_K$. Therefore, the lead electrons are described by the equilibrium Fermi function $f_K(\varepsilon_k) = [\exp((\varepsilon_k - \mu_K)/k_B T) + 1]^{-1}$. From this follows that all expectation values of the lead operators can be traced back to the expression $\langle \hat{c}_k^+ \hat{c}_{k'} \rangle = f_K(\varepsilon_k) \delta_{kk'}$ where $\delta_{kk'}$ is the Kronecker delta. As a matter of fact, we get

$$\frac{d\sigma(t)}{dt} = -\frac{i}{\hbar}[\hat{H}_{wire}, \sigma(t)] - \frac{1}{\hbar^2} \sum_{S=M,N} Tr_K \int_0^\infty dx [\hat{V}_S, [\hat{V}_S^{int}(-x), \rho(t)]]$$

(3)

where $\hat{V}_S^{int}(-x) = \exp[-\frac{i}{\hbar}(\hat{H}_{wire} + \hat{H}_{leads})x] \hat{V}_S \exp[\frac{i}{\hbar}(\hat{H}_{wire} + \hat{H}_{leads})x]$, and we used the noncrossing approximation [5]. For the evaluation of Eq. 3 it is essential to use an exact expression for the zero-order time evolution operator $\exp[-\frac{i}{\hbar}(\hat{H}_{wire} + \hat{H}_{leads})x]$.
The use of any approximation bears the danger of generating artifacts, which, for instance, may lead to a violation of fundamental equilibrium properties [6]. To do so, we first define new operators $b_f^+ = \hat{c}_{2f}^+ \hat{c}_{1f}$ and $b_f = \hat{c}_{1f}^+ \hat{c}_{2f}$ ($f=e,g$) describing charge transfer $1 \to 2$ and $2 \to 1$, respectively, in the donor-acceptor (DA) two-level system. Then the non-diagonal part of $\hat{H}_{wire}$, Eq. 2, can be rewritten in terms of operators $b_f$ only

$$\hat{H}_{wire}^{(nondiag)} = -\sum_{f=g,e}\Delta_f(b_f^+ + b_f) - \hbar J(b_e^+ b_g + b_g^+ b_e) \tag{4}$$

By expanding $\sigma$ in the many-electron eigenstates of the uncoupled sites, one obtains a $2^4 \times 2^4 = 256$ density matrix $\sigma_{\{n_{mf}\},\{n'_{mf}\}}$. Fortunately, the following consideration can be essentially simplified by using the pseudospin description based on the symmetry properties of Lie group SU(2). A two states DA system can be described by the pseudospin vector, using Pauli matrices $\hat{\sigma}_{1,2,3}$ and the unit matrix $I$ [7]. The components of the Bloch vector in the second quantization picture are given by $r_1^f = b_f^+ + b_f$, $r_2^f = i(b_f - b_f^+)$, $r_3^f = \hat{n}_{2f} - \hat{n}_{1f}$. Owing to the commutation of operators $\lambda_f = \hat{n}_{2f} + \hat{n}_{1f}$ (the electron number operator for the $f$-th DA system) and $r_i^f$, $\lambda_f$ is conserved under unitary transformations related to the diagonalization of $\hat{H}_{wire}$. Therefore, a total $2^4 \times 2^4$ space can be partitioned into nine smaller subspaces according to the values of $\lambda_f = 0,1,2$: four one-dimensional subspaces for $\lambda_f = 0,2$ (type I); four two-dimensional subspaces for $\lambda_f = 1$ and $\lambda_{f'} = 0, 2$ where $f \neq f'$ (type II); and one four-dimensional subspace for $\lambda_e = \lambda_g = 1$ (type III). One can show that

$$(r_1^f)^2 = (r_2^f)^2 = (r_3^f)^2 = \lambda_f - 2\hat{n}_{2f}\hat{n}_{1f} = \begin{cases} 0 \text{ for } \lambda_f = 0,2 \\ 1 \text{ for } \lambda_f = 1 \end{cases}, \tag{5}$$

Using Eqs. 2,4 and 5, we can write $\hat{H}_{wire}$ in terms of the Bloch vector components as follows

$$\hat{H}_{wire} = \frac{1}{2}\lambda_e(\varepsilon_{1e} + \varepsilon_{2e}) + \sum_{m=1,2} U_m N_m(N_m - 1) + \begin{cases} 0 \text{ for subspaces (I)}, \\ \frac{1}{2}r_3^f(\varepsilon_{2f} - \varepsilon_{1f}) - \Delta_f r_1^f \text{ for subspaces (II)}, \\ \frac{1}{2}\sum_{f=g,e} r_3^f(\varepsilon_{2f} - \varepsilon_{1f}) - \sum_{f=g,e}\Delta_f r_1^f - \\ -\frac{\hbar J}{2}(r_1^e r_1^g + r_2^e r_2^g) \text{ for subspace (III)} \end{cases}, \tag{6}$$

where without loss of generality we put $(\varepsilon_{1g} + \varepsilon_{2g})/2 = 0$, and the energy of the wire depends in the main on $\lambda_e$. In the following, we specify the master equation, Eq. 3, for studying two limiting cases. The first limit $U_m = 0$ describes noninteracting electrons at each sites. The second limit is the one of strong Coulomb repulsion at each site in which $U_m$ is much larger than any other energy scale of the problem. Then, only the states with at most one excess electron on the site are relevant. In both cases, a diagonal representation of the first term on the RHS of Eq. 3 is achieved by a decomposition into the eigenbasis ($\alpha, \beta$) of the many-electron wire Hamiltonian. In this basis, the fermionic interaction picture operators read

$$\langle \lambda_e, \lambda_g | \hat{c}_{nf}^{int}(-x) | \lambda_f + 1 \rangle_{\alpha\beta} = [\hat{Y}^+(\lambda_e, \lambda_g) \tilde{\chi}^+(\lambda_e, \lambda_g) \hat{c}_{nf} \tilde{\chi}(\lambda_f + 1) \hat{Y}(\lambda_f + 1)]_{\alpha\beta}$$

$$\times \exp[\frac{i}{\hbar}(E_\beta(\lambda_f + 1) - E_\alpha(\lambda_e, \lambda_g))x]$$

(7)

where $\hat{Y}(\lambda_e, \lambda_g)$ are unitary transformations related to the diagonalization of $\hat{H}_{wire}$; $\hat{\chi}(\lambda_e, \lambda_g) = \left( \{|n_{1g}, n_{2g}, n_{1e}, n_{2e}\rangle\} \right)$ is the column matrix of the many-electron eigenstates of the uncoupled sites for $n_{1e} + n_{2e} = \lambda_e$ and $n_{1g} + n_{2g} = \lambda_g$; $(\lambda_f + 1) \equiv (\lambda_e + 1, \lambda_g)$ if $f = e$ and $(\lambda_f + 1) \equiv (\lambda_e, \lambda_g + 1)$ if $f = g$; $\tilde{\chi}$ denotes the transpose matrix $\hat{\chi}$. The unitary transformations $\hat{Y}(\lambda_e, \lambda_g) = 1$ for subspaces (I). As to subspaces (II), Hamiltonian corresponding to the second line of the RHS of Eq. 6 where $\lambda_f = 1 \neq \lambda_{f'}$ can be diagonalized, using the unitary transformation

$$\begin{pmatrix} R_1^f \\ R_2^f \\ R_3^f \end{pmatrix} = \hat{T}^f \begin{pmatrix} r_1^f \\ r_2^f \\ r_3^f \end{pmatrix} \equiv \begin{pmatrix} \cos 2\vartheta_f & 0 & -\sin 2\vartheta_f \\ 0 & 1 & 0 \\ \sin 2\vartheta_f & 0 & \cos 2\vartheta_f \end{pmatrix} \begin{pmatrix} r_1^f \\ r_2^f \\ r_3^f \end{pmatrix}$$

(8)

where $\tan 2\vartheta_f = -2\Delta_f/(\varepsilon_{2f} - \varepsilon_{1f})$. The matrix elements of $\hat{T}^f$ are connected with the unitary transformations $\hat{Y}(\lambda_e, \lambda_g)$ for subspaces (II) by formula $T_{nj}^f = (1/2)Tr(\hat{\sigma}_n \hat{Y}^+ \hat{\sigma}_j \hat{Y})$ where $\hat{\sigma}_n$ and $\hat{\sigma}_j$ are Pauli matrices. The calculation of $\hat{Y}^+(1,1)$ for subspace (III) is more involved. Employing the master equation Eq. 3 and keeping for brevity only terms with $S = M$, we obtain in the wide-band limit for the steady-state condition

$$\frac{i}{\hbar}(E_\alpha - E_\beta)\sigma_{\alpha\beta} = \frac{1}{2} \sum_{nf\alpha'\beta'} \Gamma_{M,nf} \{\hat{c}_{nf,\alpha\alpha'} \sigma_{\alpha'\beta'} \hat{c}_{nf,\beta'\beta}^+ [2 - f_{K_n}(E_{\beta'} - E_\beta) - f_{K_n}(E_{\alpha'} - E_\alpha)]$$

$$+ \hat{c}_{nf,\alpha\alpha'}^+ \sigma_{\alpha'\beta'} \hat{c}_{nf,\beta'\beta}[f_{K_n}(E_\beta - E_{\beta'}) + f_{K_n}(E_\alpha - E_{\alpha'})] - \{\hat{c}_{nf,\alpha\alpha'} \hat{c}_{nf,\alpha'\beta'}^+ f_{K_n}(E_{\alpha'} - E_{\beta'})$$

$$+ \hat{c}_{nf,\alpha\alpha'}^+ \hat{c}_{nf,\alpha'\beta'}[1 - f_{K_n}(E_{\beta'} - E_{\alpha'})]\} \sigma_{\beta'\beta}$$

$$- \sigma_{\alpha\alpha'} \{\hat{c}_{nf,\alpha'\beta'} \hat{c}_{nf,\beta'\beta}^+ f_{K_n}(E_{\beta'} - E_{\alpha'}) + \hat{c}_{nf,\alpha'\beta'}^+ \hat{c}_{nf,\beta'\beta}[1 - f_{K_n}(E_{\alpha'} - E_{\beta'})]\}\}$$

(9)

$$\Gamma_{M,nf} = \frac{2\pi}{\hbar} \sum_{k \in K_n} |V_{nf,k}^{(MK_n)}|^2 \delta(\varepsilon_k - \varepsilon_{nf})$$

where                                                                                          .

## Calculation of Current

The current through the dashed line (see Fig.1) is given by $\hat{I} = e\frac{d}{dt}\hat{N} = \frac{ie}{\hbar}[\hat{H},\hat{N}]$ where $\hat{N} = \sum_{k \in L} \hat{c}_k^+ \hat{c}_k + \hat{n}_{1g} + \hat{n}_{1e}$ is the operator of the electron number on the left from the dashed line. Calculating the commutator on the RHS of the last equation, we get $\hat{I} = \frac{ie}{\hbar} \sum_{f=g,e} \Delta_f (b_f - b_f^+) = \frac{e}{\hbar} \sum_{f=g,e} \Delta_f r_2^f$. Using Eq. 6, we obtain

$$\hat{I} = \frac{e}{\hbar} \sum_{f=g,e} \Delta_f r_2^f (\lambda_f = 1) \qquad (10)$$

Obviously $\lambda_f = 1$ in Eq. 10 is another way of saying that the current in channel $f$ exists only for the case of one of states $\{f\}$ is occupied and another one of $\{f\}$ is unoccupied.

## Strong Coulomb Repulsion at Sites

In the limit of strong Coulomb repulsion, $U_m$ is assumed to be so large that at most one excess electron resides on each site. Thus, the available Hilbert space for uncoupled sites is reduced to three states $\hat{\chi}(0,0)$, $\hat{\chi}(2,0)$ and $\hat{\chi}(0,2)$ for subspaces I; two states $\hat{\chi}(1,0)$ and $\hat{\chi}(0,1)$ for subspaces II; and the state $\hat{\chi}(1,1) = \begin{pmatrix} |1_{1g},0_{2g},0_{1e},1_{2e}\rangle \\ |0_{1g},1_{2g},1_{1e},0_{2e}\rangle \end{pmatrix}$ for subspace III, which in this case becomes two-dimensional one. Consider subspaces (II). The matrix $\hat{T}^f$, Eq. 8, with matrix elements $T_{nj}^f = (1/2)Tr[\hat{\sigma}_n \hat{Y}^+(\lambda_f = 1; \lambda_{f'} = 0) \hat{\sigma}_j \hat{Y}(\lambda_f = 1; \lambda_{f'} = 0)]$ describes a rotation by mixing angle $2\vartheta_f$ around axis "$y$". $\hat{Y}(\lambda_f = 1; \lambda_{f'} = 0)$ is an unitary operator defined by

$$\hat{Y}^+(\lambda_f = 1; \lambda_{f'} = 0) = \begin{pmatrix} \cos\vartheta_f & \sin\vartheta_f \\ -\sin\vartheta_f & \cos\vartheta_f \end{pmatrix} \qquad (11)$$

which enables us to obtain eigenstates

$$\begin{pmatrix} \Phi_+(\lambda_f = 1; \lambda_{f'} = 0) \\ \Phi_-(\lambda_f = 1; \lambda_{f'} = 0) \end{pmatrix} = \hat{Y}^+(\lambda_f = 1; \lambda_{f'} = 0) \hat{\chi}(\lambda_f = 1; \lambda_{f'} = 0) \qquad (12)$$

and eigenvalues

$$E_\pm(\lambda_f = 1; \lambda_{f'} = 0) = \frac{1}{2}[\lambda_e(\varepsilon_{1e} + \varepsilon_{2e}) + (\varepsilon_{2f} - \varepsilon_{1f}) \pm \sqrt{(\varepsilon_{2f} - \varepsilon_{1f})^2 + 4\Delta_f^2}]) \qquad (13)$$

for subspaces (II). As to subspace III, operator $\hat{Y}^+(1,1)$ for $\varepsilon_{ng} = 0$, $\varepsilon_{ne} = \varepsilon_e$ and $\Delta_g = 0$ is

reduced to

$$\hat{Y}^+(1,1) = \frac{1}{\sqrt{2}} \begin{pmatrix} 1 & 1 \\ -1 & 1 \end{pmatrix}$$

(14)

It enables us to obtain the corresponding eigenstates $\hat{\Phi}(1,1) = Y^+(1,1)\hat{\chi}(1,1)$ and eigenvalues $E_{1,2} = \varepsilon_e \mp J\hbar$. Using Eqs. 10 and 14 and taking the expectation value of current, we get for $\Delta_g = 0$

$$\langle I \rangle = \frac{2e}{\hbar} \Delta_e \operatorname{Im} \sigma_{-+}(1,0)$$

(15)

## Current from Electronic Excitations in the Wire

Recently Galperin, Nitzan and Ratner [8] predicted the existence of non-Landauer current induced by the electron-hole excitations in the leads. Here we show that the non-Landauer current is induced also by the exciton type excitations in the wire itself. Consider a strong bias limit in the Coulomb blockade case, where $\mu_L > \varepsilon_e$ and $\mu_R < \varepsilon_g$, and the states $\varepsilon_f$ are positioned rather far ( $\gg k_B T, |J|, |\Delta_e|$ ) from the Fermi levels of both leads so that $f_L(\varepsilon) = 1$ and $f_R(\varepsilon) = 0$ on the RHS of Eq. 9. The Landauer current in the case under consideration ( $\Delta_g = 0$ ) occurs in channel "e" when it is isolated from channel "g" that is realized for $\Gamma_{M,1g} = \Gamma_{M,2g} = 0$ and $\lambda_g = 0$. The latter equality enables us to avoid blocking the current in channel "e" due to strong Coulomb repulsion at sites. Indeed, the Landauer current does not exist for $\Gamma_{M,1e} = \Gamma_{M,2e} = 0$ even when $\Gamma_{M,1g}, \Gamma_{M,2g} \neq 0$, since $\Delta_g = 0$. In contrast, the solution of Eq. 9 in the rotating-wave approximation (RWA) [3] gives for this case

$$\sigma_{-+}(1,0) = \frac{i\hbar}{-8\Delta_e} \{\Gamma_{M,1g} Tr\sigma(1,0) + \Gamma_{M,2g} Tr\sigma(1,1)\}$$

(16)

and $Tr\sigma(1,0) = (\Gamma_{M,2g}/\Gamma_{M,1g}) Tr\sigma(1,1)$. Then using the normalization condition $Tr\sigma(1,0) + Tr\sigma(1,1) = 1$ and Eq. 15, we obtain

$$\langle I \rangle_{RWA} = -\frac{e}{2} \frac{\Gamma_{M,2g} \Gamma_{M,1g}}{\Gamma_{M,2g} + \Gamma_{M,1g}}$$

(17)

The last equation describes a non-Landauer current caused by the electron transfer occurring in different channels: the intersite transfer in the bridge takes place in channel "e", and the bridge-metals charge transfer occurs in channel "g". The inter-channel mixing is induced by the energy-transfer term $\sim J$ (see Fig.2). Consider a cycle corresponding to the charge transfer of $e$. Initially electrons populate states $|1g\rangle$ (due to coupling to the left lead, $\Gamma_{M,1g} \neq 0$ ) and $|2e\rangle$ ( $\Gamma_{M,2e} = 0$ ) of the sites (Fig.2a). The energy transfer induces the following transitions: $|2e\rangle \to |2g\rangle$ and $|1g\rangle \to |1e\rangle$ (vertical arrows). The system arrives to the state shown in Fig.2b. Due to the coupling to the right lead ($\Gamma_{M,2g} \neq 0$), electron from state $|2g\rangle$ moves into the right lead (the horizontal arrow), and after this the system is described by Fig.2c. Due to releasing the right site and the hopping matrix element $\Delta_e$, electron from state $|1e\rangle$ passes into state $|2e\rangle$ (the upper horizontal arrow), releasing the left site. Then an electron from the left lead moves into state $|1g\rangle$ ($\Gamma_{M,1g} \neq 0$, the lower horizontal arrow), and the system returns into the initial state described by Fig.2a.

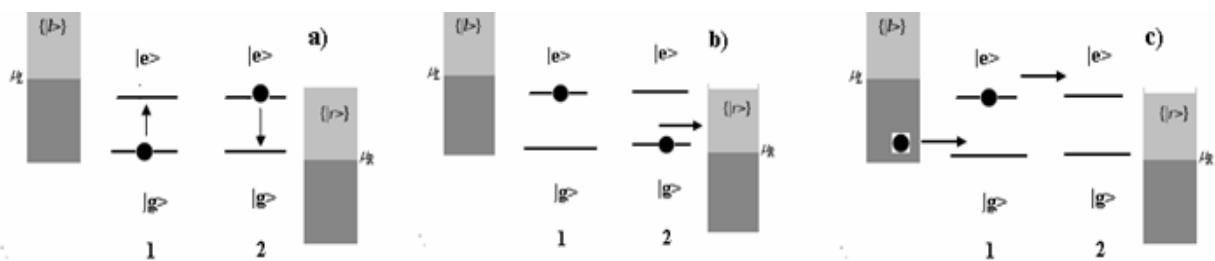

**FIGURE 2.** Different stages of the energy-transfer induced current. a) energy transfer, $\sigma(1,1)\neq 0$. b) the charge transfer to the right lead. c) the intersite charge transfer, $\sigma(1,0)\neq 0$; the charge transfer from the left lead.

## OPTICAL CONTROL OF CURRENT WITH CHIRPED PULSES

In the second part of the paper we describe a theory for light-induced current by strong optical pulses in unbiased molecular tunneling junctions as a special case where the second condition for finite current (see Introduction) - a biased junction - is not fulfilled. We consider a class of molecules characterized by strong charge-transfer transitions into their first excited state [9]. We have proposed a novel control mechanism by which the charge flow is enhanced by chirped pulses. For linear chirp and when the energy transfer between the molecule and electron-hole excitations in the leads is absent, this control model can be reduced to the Landau-Zener transition to a decaying level. The details can be found in Ref. [5].

## ACKNOWLEDGMENTS

This work was supported by the German-Israeli Fund (P.H., S.K. and A.N.), ISF (B.F. and A.N.) and the NIM excellence program (B.F.).

## REFERENCES

1. *Proc. Natl. Acad. Sci. U.S.A.* **102**, 8800 (2005), special issue on molecular electronics, edited by C. Joachim and M. A. Ratner.
2. S. Kohler, J. Lehmann, and P. Hanggi, *Phys. Reports* **406**, 379-443 (2005).
3. F. J. Kaiser, M. Strass, S. Kohler and P. Hanggi, *Chem. Phys.* **322**, 193-199 (2006).
4. A. S. Davydov, "Theory of Molecular Excitons", New York, Plenum, 1971.
5. B. D. Fainberg, M. Jouravlev, and A. Nitzan, *Phys. Rev. B* **76**, 245329 (2007).
6. T. Novotny, *Europhys. Lett.* **59**, 648-654 (2002).
7. F. T. Hioe and J. H. Eberly, *Phys. Rev. Lett.* **47**, 838-841 (1981).
8. M. Galperin, A. Nitzan and M. A. Ratner, *Phys. Rev. Lett.* **96**, 166803 (2006).
9. M. Galperin and A. Nitzan, *Phys. Rev. Lett.* **95**, 206802 (2005).